\documentclass{article}

\usepackage{amssymb}

\usepackage[figuresright]{rotating}

\begin{document}





\title{ Background Fluctuations in Heavy Ion Jet Reconstruction}


\author{P.~M.~Jacobs \thanks{pmjacobs@lbl.gov} \thanks{for the STAR Collaboration} 
\thanks{Proceedings of Hard Probes 2010, Eilat, Israel, Oct. 10-15 2010.}\\
Lawrence Berkeley National Laboratory, Berkeley CA, USA 94720}


\maketitle


\begin{abstract}
We present a new study by the STAR Collaboration of background fluctuations in jet reconstruction in heavy ion collisions.
\end{abstract}


\newcommand{\pT}{\ensuremath{p_T}}
\newcommand{\kT}{\ensuremath{k_T}}
\newcommand{\antikT}{anti-\ensuremath{k_T}}
\newcommand{\AuAu}{\ensuremath{Au+Au}}
\newcommand{\sNN}{\ensuremath{\sqrt{s_{NN}}}}
\newcommand{\pp}{\ensuremath{p+p}}
\newcommand{\s}{\ensuremath{\sqrt{s}}}

\newcommand{\pTembed}{\ensuremath{\pT^{embed}}}
\newcommand{\pTreco}{\ensuremath{\pT^{reco}}}
\newcommand{\Areco}{\ensuremath{A^{reco}}}
\newcommand{\dpT}{\ensuremath{\delta\pT}}


Jets are fundamental objects in QCD, and are ubiquitous in high energy collisions of all kinds. In heavy ion collisions we utilize jets to probe the hot medium. While the rate of initial hard (high $Q^2$) scattering is not altered by the presence of medium, the interaction of the scattered partons with the medium induces changes to the internal jet structure, and may alter the jet direction. It is the modification of internal jet structure and of inter-jet correlations that are the observables of jet quenching~\cite{RHICHadronSupp}. 

Initial studies of jet quenching at RHIC, utilizing high \pT\ hadrons and their correlations as jet surrogates, have found marked suppression effects that establish jet quenching experimentally \cite{RHICHadronSupp}. Extensive effort is underway to understand these measurements theoretically \cite{RHICHadronSupp,Marco} and to use them to quantify medium transport properties \cite{Nestor}. While such hadronic measurements have been valuable to establish the phenomenon of jet quenching and provide initial quantitative constraints on medium properties, their scope is limited by the biases inherent in leading-hadron observables~\cite{BaierBias}, i.e. the observed high \pT\ hadron population is dominated by remnants of those jets that have interacted {\it least} with the medium. Full study of the dynamics of jet quenching requires unbiased measurement of jets that have undergone significant interaction in matter, either via jets recoiling from a direct photon or Z boson, or via unbiased jet reconstruction that recovers the energy ``flow" of the scattered parton, independent of how it fragments. 

The measurement of jets in high energy heavy ion collisions is more complex than in elementary collisions, due to the large fluctuations in the underlying background. The first measurement required in such a program is the inclusive jet cross section, which quantifies the degree to which unbiased jet reconstruction has been achieved  in practice. Preliminary STAR measurements of the inclusive jet cross section at mid-rapidity in central \AuAu\ collisions at \sNN=200 GeV have been reported previously \cite{SevilMateusz}, based on the \kT\ and \antikT\cite{antikT,FastJet} algorithms and the FastJet background correction scheme \cite{FastJetHI,Soyez}. STAR measures jets in its central barrel ($\left|\eta\right|<1.0$) utilizing charged particle tracking and electromagnetic calorimetry \cite{STARppJet}. Underlying event fluctuations distort the inclusive jet distribution significantly, and the corresponding correction is by far the largest source of systematic uncertainty in this measurement. 

The FastJet-based background correction to the inclusive cross section measurement proceeds in two steps:

\begin{enumerate}
\item Event-wise background estimate: reconstruct all jets in the experimental acceptance using the \kT\ algorithm with $R=0.4$ \cite{FastJet,FastJetHI} and determine the median energy density $\rho=median\{\pT^i/A^i\}$, where $\pT^i$ and $A^i$ are the \pT\ and area of the $i$th jet in the event, $1\le{i}\le{N_{jet}}$. Corrected jet \pT\ is then $\pT^{i,corr}=p_T^i-\rho\cdot{A_i}$. This accounts on average for the underlying event background, but not residual fluctuations.
\item Deconvolution of fluctuations: estimate inclusive spectrum distortion due to fluctuations; express in terms of bin migration and correct by regularized matrix inversion. 
\end{enumerate}

In \cite{SevilMateusz}, the spectrum distortion in step (2) was estimated by generating PYTHIA jets with a realistic \pT\ spectrum for \pp\ collisions at \s=200 GeV, injecting (or "embedding") them into real STAR central \AuAu\ events, and reconstructing the hybrid events with the same procedure as used for the data analysis. The resulting spectrum distortion was parameterized as the convolution of a Gaussian function with $\sigma=6.8$ GeV and systematic uncertainty $\pm1$ GeV. While this is a well-controlled procedure, it does not address two important issues: (i) sensitivity of the correction to the fragmentation pattern of the signal jets, of particular concern since  fragmentation of quenched jets is not modeled by PYTHIA, and (ii) non-Gaussian response, especially a non-Gaussian tail towards positive fluctuations that can distort the spectrum over a wide \pT\ range.

In these proceedings we present a new study that addresses these issues. This study is likewise based on embedding of a simulated jet population into real STAR central \AuAu\ events, but with a wider spectrum of jet models than the previous study: single hadrons,  PYTHIA jets, and Q-PYTHIA jets to model quenching effects \cite{QPYTHIA}. Embedding is carried out at the level of charged particle tracks and calorimeter energy deposition, without accounting at present for instrumental effects in order to isolate the irresolution due solely to background fluctuations. For the background population we use 8M STAR \AuAu\ central (0-10\%) events, from an event sample recorded with a minimum bias trigger in the 2007 RHIC run.

We embed an object of known \pT=\pTembed\ and apply jet reconstruction to the hybrid event, utilizing the FastJet \antikT\ algorithm with $R=0.4$. We find all reconstructed jets containing embedded charged tracks and calorimeter towers, and identify the matched jet as that reconstructed jet containing over 50\% of the total \pT\ of the embedded jet. The matching efficiency is found to be high (over 90\% at low \pT, over 95\% for \pT$>$10 GeV). For embedded \pTembed\ and matched jet \pTreco, we quantify the response of the hybrid system to the embedded jet via:

\begin{equation} 
\dpT = \pTreco - \rho\cdot\Areco - \pTembed,
\label{eq:dpT}
\end{equation}

\noindent
where \Areco\ is the area of the matched reconstructed jet and $\rho$ is determined prior to the embedding step. This definition is identical to Eq. (1) in \cite{FastJetHI}. The normalized distribution of \dpT\  is the probability distribution to find jet energy (after event-wise background correction) $\pT^{corr}=\pT^{true}+\dpT$. If there were no background fluctuations, \dpT\ would be a delta function at zero.


\begin{figure}[htbp]
   \centering
   \includegraphics[width=0.49\textwidth]{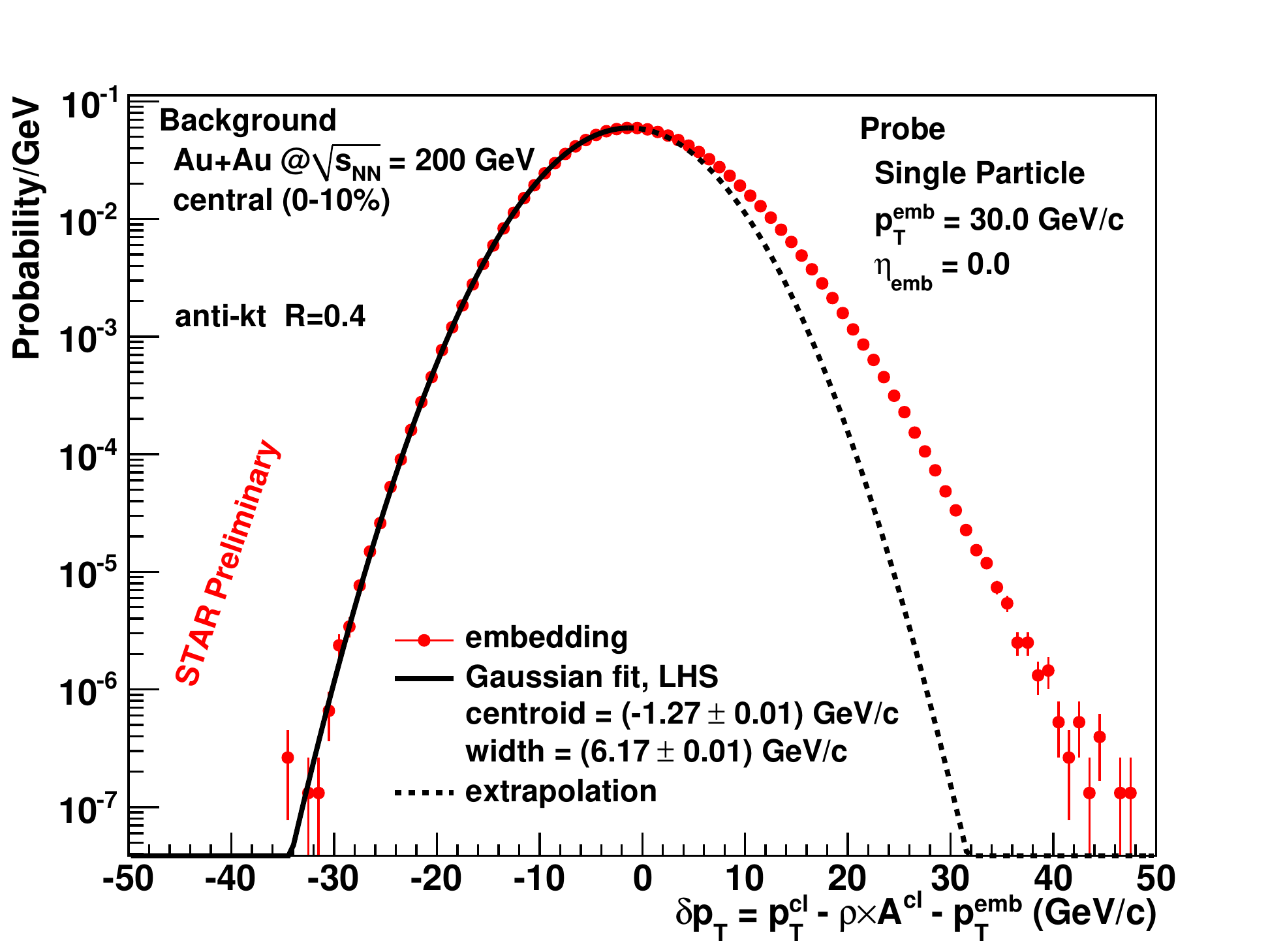} 
   \includegraphics[width=0.49\textwidth]{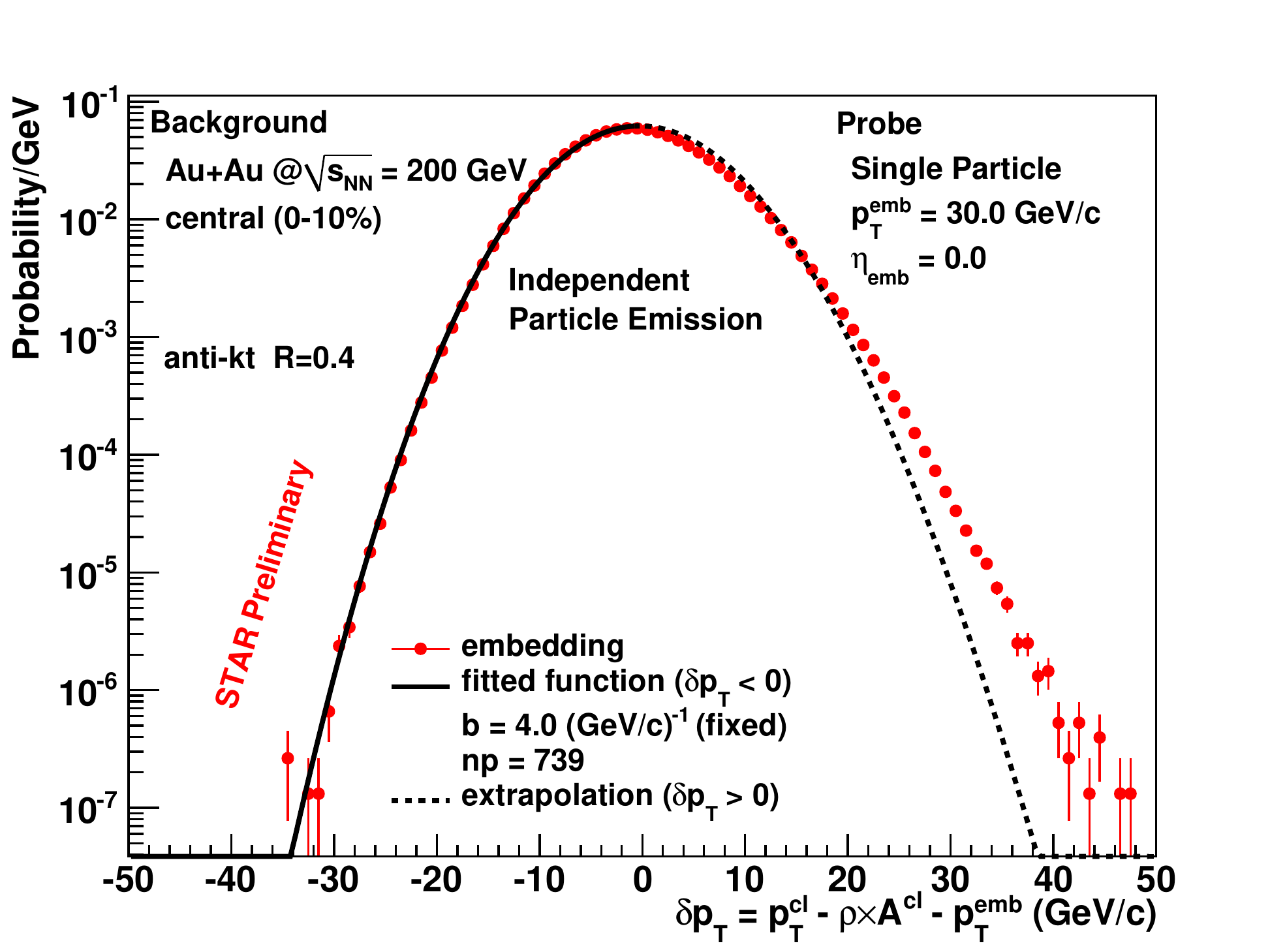} 
   \caption{Single particle embedding: distribution of \dpT (Eq. \ref{eq:dpT}) for a ``jet"  corresponding to a single pion with \pT=30 GeV embedded into central \AuAu\ events. The same \dpT\ distribution is shown in both panels. Left panel shows fit with Gaussian function, right panel shows fit with Gamma function modeling independent particle emission\cite{Tannenbaum}. See text for details.
   \label{fig:SPE}}
\end{figure}

We first consider the embedding of a single hard pion. While this is an unlikely fragmentation pattern for a jet, its simplicity aids in interpretation and we find that single particle embedding captures many of the features of the system response. We will later compare the single particle response with that of more physical jet fragmentation patterns, in order to isolate the specific role of fragmentation. Fig.~\ref{fig:SPE} shows the normalized distribution of \dpT\ for a single pion with \pT=30 GeV, $\eta=0.0$ and $\phi=0$ (i.e. far from STAR acceptance boundaries), embedded successively in 8M different STAR central collision events. It is the ensemble-averaged response to this particular choice of jet probe. The probability to reconstruct a 30 GeV single-particle jet with greater than 20 GeV in excess energy (after event-wise background correction) is $\sim10^{-3}$. In order to characterize the distribution we fit with two analytic functions:

\begin{itemize}
\item Fig. \ref{fig:SPE}, left panel, shows a Gaussian function fit to the region of \dpT\ less than the median of the distribution. The fit gives a centroid -1.27 GeV and width 6.17 GeV. The region $\dpT<0$ is well modeled as a Gaussian but the region  $\dpT>0$ has significant excess relative to the fit, which can strongly influence the inclusive spectrum measurement. A Gaussian function is not an accurate model of the distribution of  fluctuations \cite{FastJetHI}.
\item Fig. \ref{fig:SPE}, right panel, shows a fit to the region $\dpT<0$ of an analytic function describing the transverse energy distribution into finite acceptance of multiple independent particle emission sources, each emitting an exponentially falling spectrum 
\cite{Tannenbaum}. The spectrum has fixed $\left<\pT\right>=0.5$ GeV/c, corresponding roughly to RHIC conditions, and the fit finds about 370 independent ``sources". The fit function is non-Gaussian at large \dpT. This simple model of uncorrelated particle emission can account for the bulk of the irresolution due to background fluctuations.
\end{itemize} 


\begin{figure}[htbp]
   \centering
   \includegraphics[width=0.40\textwidth]{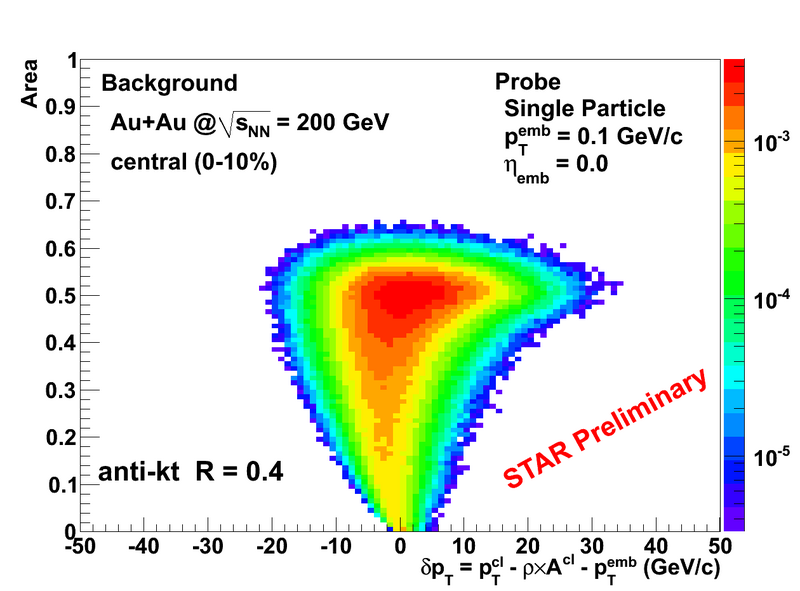} 
   \includegraphics[width=0.40\textwidth]{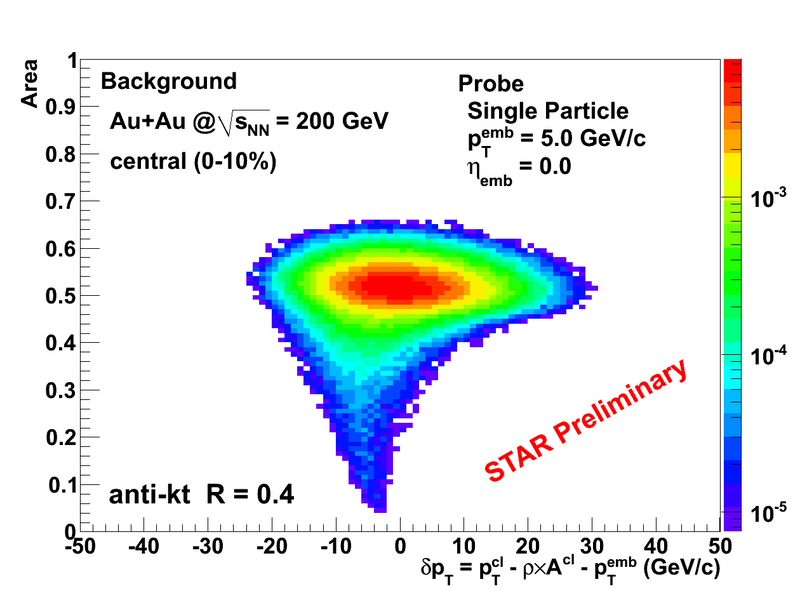} 
   \includegraphics[width=0.40\textwidth]{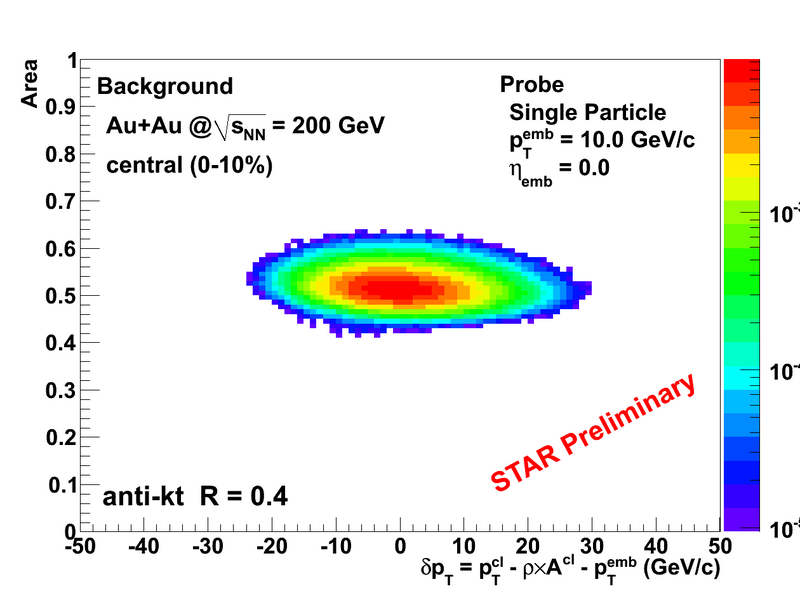} 
   \includegraphics[width=0.40\textwidth]{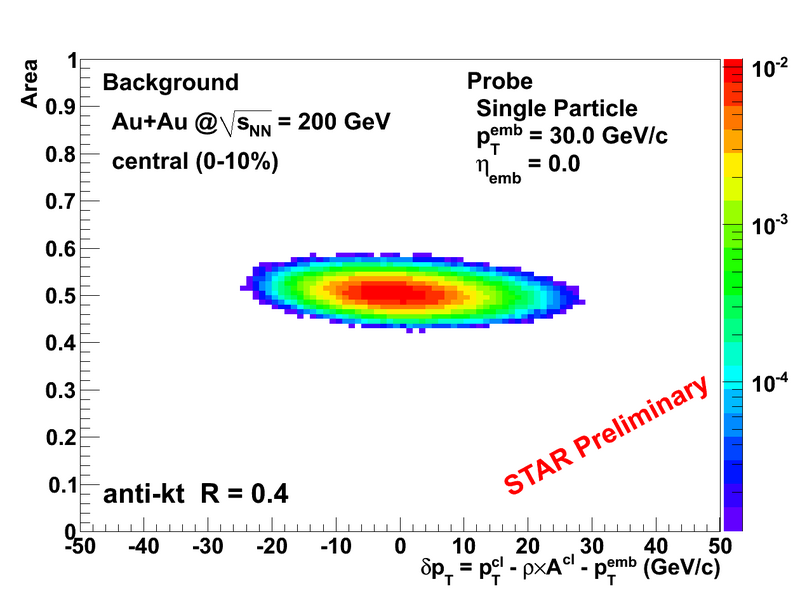} 
   \caption{Jet area vs. \dpT\ for embedded single pions with \pT=0.1 (upper left), 5.0 (upper right), 10.0 (lower left) and 30.0 (lower right) GeV.}
   \label{fig:AreavspT}
\end{figure}

\begin{figure}[htbp]
   \centering
 \includegraphics[width=0.39\textwidth]{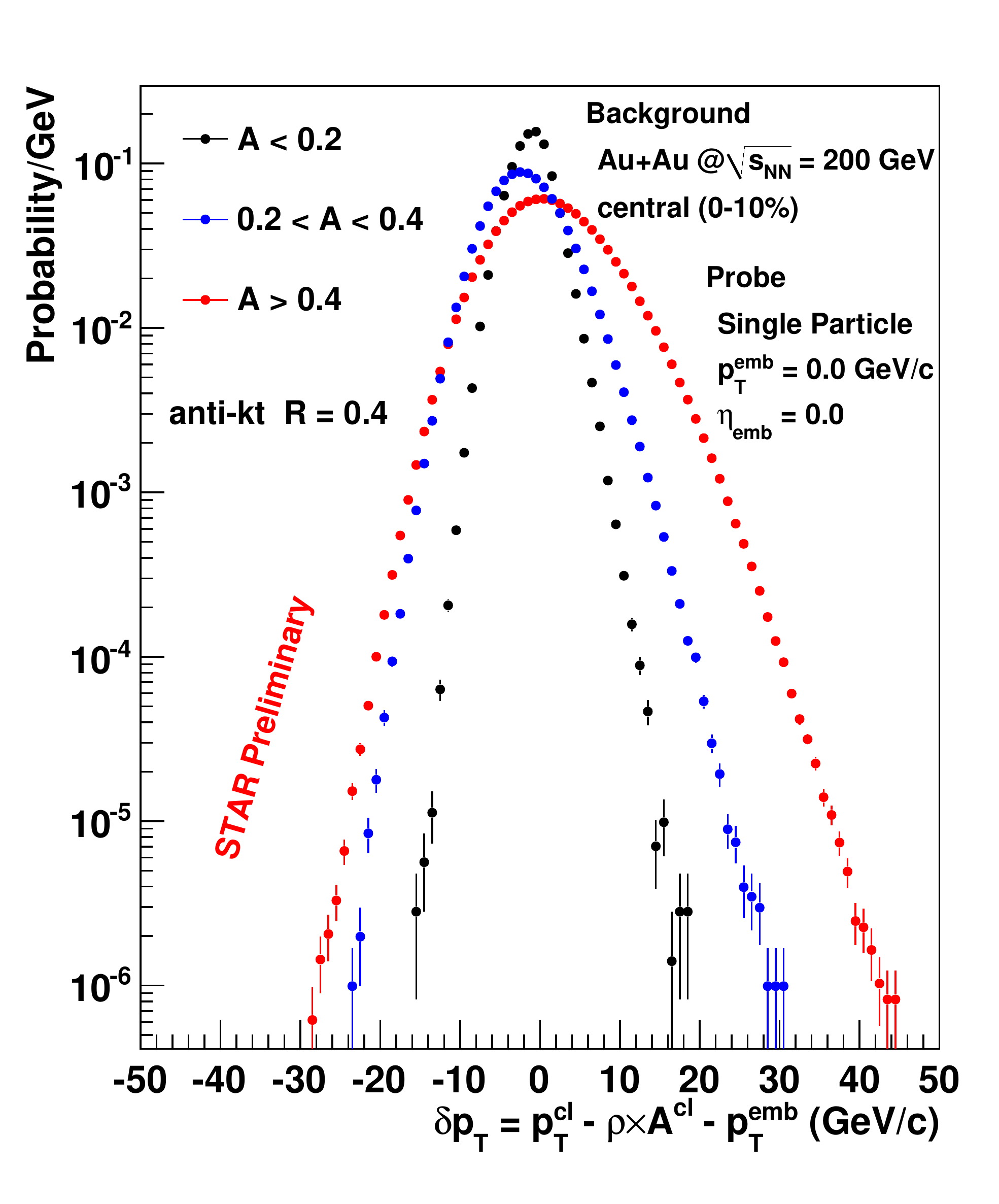} 
 \includegraphics[width=0.58\textwidth]{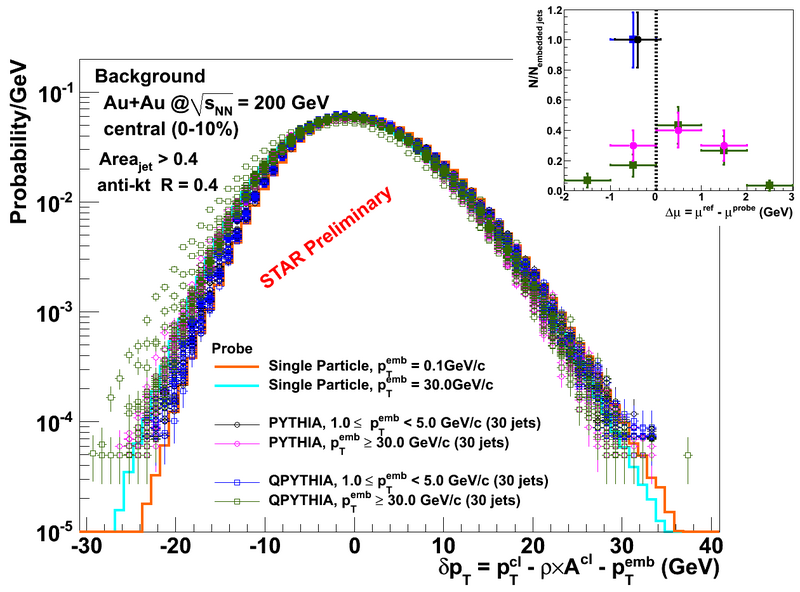} 
   \caption{\dpT\ distribution systematics. Left: for different jet areas, for probe with \pT=0.1 GeV. Right: overlay of \dpT\ for single pions, PYTHIA and Q-PYTHIA jets, of varying \pT.
   \label{fig:dpTsystematics}}
\end{figure}

Background fluctations are generally expected to be correlated with jet area; for background due to uncorrelated emission, $\sigma_{fluct}\sim\sqrt{A}$ \cite{FastJet,FastJetHI}. Fig. \ref{fig:AreavspT} shows the distribution of measured jet area \cite{FastJet,FastJetHI} vs. \dpT, for embedded single pion jets with a broad range of \pT: 0.1, 5, 10 and 30 GeV. At large \pT, there is a narrow area distribution around the expected value $\pi{R^2}\simeq0.5$ for $R=0.4$, with little dependence of the \dpT\ distribution on \pT. At lower \pT\ there is a significant population of lower area jets, due to the response of the \antikT\ algorithm in dense background for jets lacking an energetic core. The \dpT\ distribution is narrower for smaller jet area, as seen in Fig. \ref{fig:dpTsystematics}, left panel. The deconvolution of background fluctuations from the inclusive jet spectrum must accurately take into account the \pT-dependence of the jet area distribution, and consequently the \pT-dependence of fluctuations (i.e. of \dpT); this is the subject of forthcoming work.


Finally, we investigate the dependence of \dpT\ on fragmentation pattern and \pT. Fig. \ref{fig:dpTsystematics}, right panel, shows the overlay of multiple \dpT\ distributions for single particle jets and for jets generated by PYTHIA and Q-PYTHIA ($\hat{q}$=5 GeV$^2$/fm). For both PYTHIA and Q-PYTHIA, 30 different jets within each of the intervals $1<\pT<5$ GeV and $\pT>30$ GeV were embedded. In order to compare their shapes directly, the distributions were aligned horizontally by fitting a Gaussian function to \dpT$<0$ and aligning the centroids by shifting relative to one reference distribution. The shifts are shown in the insert and are typically smaller in magnitude than 1 GeV. The overlay shows that the \dpT\ distribution is to a large extent universal, within a factor $\sim2$ at \dpT=30 GeV. Two energetic  jets from the Q-PYTHIA population do not follow the bulk of the distribution for $\dpT<-10$ GeV, though this will have little impact on the smearing of the inclusive spectrum. 

Insensitivity of the \dpT\ distribution to fragmentation pattern is crucial for well-controlled analysis of quenched jets, whose fragmentation is {\it a priori} unknown. Fig. \ref{fig:dpTsystematics}, right panel, indicates that reconstruction in a dense background environment using the \antikT\ algorithm \cite{antikT} is indeed robust against variations in fragmentation. Further quantification of this observation, and its application to deconvolution of the measured inclusive jet spectrum in central \AuAu\ collisions, is in progress.





\bibliographystyle{elsarticle-num}

\begin{thebibliography}{00}

\bibitem{RHICHadronSupp} A. Majumder and M. van Leeuwen, arXiv:1002.2206.

\bibitem{Marco} M. van Leeuwen, these proceedings.

\bibitem{Nestor} N. Armesto {\it et al.}, J. Phys. {\bf G37}, 025104 (2010).

\bibitem{BaierBias} R. Baier, Nucl. Phys. {\bf A715}, 209 (2003).

\bibitem{SevilMateusz} S. Salur {\it et al.}, Eur.~Phys.~J. {\bf C61} 761 (2009); M. Ploskon {\it et al.}, Nucl.~Phys.~{\bf A830} 255C (2009).

\bibitem{antikT} M. Cacciari, G. Salam and G. Soyez, JHEP {\bf 0804}, 063 (2008).

\bibitem{FastJet} M. Cacciari and G. Salam, Phys.~Lett.~B641, 57 (2006); M. Cacciari, G. Salam and G. Soyez, JHEP {\bf 0804} 005 (2008); http://fastjet.fr/.

\bibitem{FastJetHI} M. Cacciari, J. Rojo, G. Salam and G. Soyez, arXiv:1010.1759 (hep-ph).

\bibitem{Soyez} G. Soyez, these proceedings.

\bibitem{STARppJet} B.I. Abelev {\it et al.} (STAR), Phys. Rev. Lett. {\bf 97}, 252001(2006).

\bibitem{QPYTHIA} N. Armesto {\it et al.}, Eur.~Phys.~J. {\bf C63}, 679 (2009).

\bibitem{Tannenbaum} M. Tannenbaum, Phys. Lett. {\bf B498}, 29 (2001).

\end{thebibliography}



\end{document}